# Overcrowding drives the unjamming transition of gap-free monolayers

Tao Su[*] and Ganhui Lan[*†]

**Abstract**

Collective cell motility plays central roles in various biological phenomena such as inflammatory response, wound healing, cancer metastasis and embryogenesis. These are biological demonstrations of the unjamming transition. However, contradictory to the typical density-driven jamming processes in particulate assemblies, cellular systems often get unjammed in highly packed, sometimes overcrowding tissue environments. In this work, we report that overcrowding can unjam gap-free monolayers through increasing isotropic compression. The transition boundary is determined by the isotropic compression and the cell-cell adhesion. We explicitly construct the free energy landscape for the T1 topological transition during monolayer rearrangement, and find that it evolves from single-barrier shape to double-barrier shape upon completion of the unjamming process. Our analyses reveal that the overcrowding and adhesion induced unjamming transition reflects the mechanical yielding of the highly deformable monolayer, which differs from those caused by loosing up a packed particulate assembly.

[*] Department of Physics, George Washington University, Washington DC USA
[†] Address correspondence to: Ganhui Lan (glan@gwu.edu)

Unjamming transition happens ubiquitously in crowded many-body systems[1-5]. It describes the change of the overall mobility within the system and is related to the system's rigidity transition from a solid-like phase to a liquid-like phase[6]. In non-living particulate assemblies, such as granular mixtures[7] and colloidal suspensions[6, 8], such transition is reflected by the dramatically changed diffusibility of the embedded particles[8, 9]. In living systems, such as embryo[10, 11] and epithelial monolayer[12-14], the transition can be characterized by the diminishing of the energy barrier for cells to switch neighbors via the T1 topological transition[15-17] while moving within the tissue[13, 18, 19], which is deeply connected to many important physiological processes, including wound healing[20-23], embryo development[10, 11, 14], cancer progression[18, 24, 25] and immune responses[26].

Experimental and theoretical studies suggested that the unjamming transitions in non-living and living systems share similar properties such as heterogeneity[27-29]. Both types of systems exhibit collective motility while approaching the jamming onset, and then reach immobile solid-like phase with relatively stable yet disordered morphologies[28-31]. Unjamming transition is largely determined by the systems' physical properties[32, 33], including the interactions between different units, the active forces/motions of the units and the system's density. Models using self-propelled objects revealed that the actively moving particles tend to align their motions through short-range interactions when density approaching a critical value[30, 34-36]. Similarly, in moving cellular systems, experiments have showed that increasing density causes the emergence of localized groups of motile cells moving coordinately[5, 27, 29-31]. The size of these local groups increases but their moving velocity decreases when cell density increases[27, 37, 38]. These observations are analogous to the jammed motion in the Time Square

Ball Drop events: individuals in the crowd can only move slowly with their neighbors (i.e. coordinated motile group) or cannot move at all (i.e. jammed).

The unjamming phenomena in tissue-like cellular systems are somewhat unique. Here, the gaps between cells are often negligible[30, 37, 39, 40] with packing fraction $\phi$ at 1. It has been recently discovered that while keeping unity $\phi$, such gap-free cellular monolayer can switch between the solid-like and liquid-like states by adjusting a *target shape index* $p_0$ that is determined by the strengths of the adhesive and tensional interactions between cells[41, 42]. And the unjamming transition occurs at a critical index value around $p_0^* = 3.81$ [41]. This discovery suggests that crowdedly packed cellular systems may undergo unique unjamming transitions to regulate their collective motility.

Intensive attentions have been devoted to the field of unjamming related collective cell migration. Many successful vertex models have been applied to investigate the morphologies of crowdedly packed cellular monolayers and their relationship to the systems' motility[13, 14, 43-49]. Current theoretical studies have been converging to a general picture that the relevant intracellular molecules control cell packing by changing the force balances among the cells, re-adjusting the cells' geometries, and eventually contributing to collective movements in crowdedly packed many-cell systems[5, 13, 21, 29, 44, 47, 50-52]. Given the scales of the relevant forces and lengths, most of these approaches focused on the ground state of the monolayers where the mechanical energy is minimized[44, 53].

Here, we combine the vertex model with a Monte Carlo stimulation scheme to investigate the monolayer's collective behaviors when cell number changes, especially increases, under the gap-free constraint. Counterintuitively, we find that increased cell number in a confined 2-dimensional space (i.e. higher cell density) can unjam the gap-free monolayer to reach liquid-like phase. By constructing the free energy landscapes for the T1 topological transition, we discover that, instead of simply diminishing, the free energy landscape evolves from single-barrier shape to double-barrier shape during unjamming, which may be responsible for the dynamics of the collective cell rearrangement within the monolayer. Our analyses also reveal a morphological transition of cells' geometrical properties from a distributed state to a disordered state, coinciding with the unjamming transition. These results suggest an alternative mechanism that cells may spontaneously gain collective mobility simply through cell growth and division in confined tissue and engineered environments.

**Results**

**The vertex-based stochastic modeling scheme**. We use the well-documented vertex model to mesh the 2-dimensional monolayer into polygons for approximation of individual cells[13, 43, 45, 47, 48]. Each edge represents the contacting membranes of two adjacent cells. In the crowdedly packed monolayer, because there are no gaps between cells, the efficient Voronoi tessellation algorithm can be applied to conduct the meshing[41, 54-56]. Once the monolayers are meshed into individual cells, a simplified form of energy function has been widely used to calculate the mechanical energy of the cellular system[13, 41, 44, 45, 56]:

$$U = \sum_{i=1}^{N} U_i = \sum_{i=1}^{N} [\alpha(A_i - A_0)^2 + \beta P_i^2 + \gamma P_i] \qquad (1)$$

where $N$ is the total number of cells in the monolayer; $A_i$ and $P_i$ are the $i$th cell's projected area and perimeter on the monolayer plane, respectively; $A_0$ is the preferred cell area; $\alpha$, $\beta$ and $\gamma$ represent the mechanical compliances of the area compression, perimeter contraction and linear tension, respectively[40, 57]. In particular, $\alpha$ and $\beta$ are positive, whereas $\gamma$ can be either positive or negative, depending on the relative strength of tensional (positive) and adhesive (negative) interactions along the common edges of adjacent cells. Equation (1) nicely summarizes the most relevant mechanical contributions within a cellular monolayer with homogeneous cell properties. Because the scales of forces and lengths are respectively in the order of nano-Newton and micrometer, the mechanical energy computed from Equation (1) is much larger than thermal energy, and the monolayer behaviors have therefore been mostly studied at the ground state (i.e. the minimum mechanical energy state)[44, 53].

But cellular systems are noisy and intrinsically dissipative and the ground state configurations do not correspond to realistic tissue morphologies[13]. To account for the entropic effect of the monolayer at the cellular level, we consider the uncoordinated active mechanical agitations $\boldsymbol{f_t}$ that are also able to deform the cells. Here, we assume that the time average $\langle \boldsymbol{f_t} \rangle_t = 0$; however, the work, $W_f$, done by these agitations is not *zero* and can contribute to the morphological dynamics of the monolayer by effectively decreasing the cells' mechanical energy:

$$U' = U - W_f = \left(1 - \frac{W_f}{U}\right)U = \varepsilon U \qquad (2)$$

Thus, $\boldsymbol{f_t}$ acts like the "collision event". When these uncoordinated mechanical agitations are very active, the pre-coefficient $\varepsilon$ can be very small, meaning that the cells are subjected to uncoordinated perturbations that may significantly influence their morphological dynamics. We use "uncoordinated" instead of "random" to describe those active mechanical agitations because many of these events may still under the regulation of other biological pathways that are not directly coordinated with the cell-cell contact or cell motility[19].

We rescale the parameters and rewrite Equation (1):

$$U' = \sum_{i=1}^{N} U'_i = \sum_{i=1}^{N} \alpha_A [(a_i - a_0)^2 + k(p_i - p_0)^2 - u_0] \qquad (3)$$

where $a_i = A_i/A_0$ and $p_i = P_i/\sqrt{A_0}$ are the nondimensionalized area and perimeter of the $i$th cell, respectively; $k = \beta/(\alpha A_0)$ is the dimensionless module measuring the relative elastic contributions from perimeter and area; $\alpha_A = \varepsilon \alpha A_0^2$ is the overall energy coefficient of the cell and reflects the effective temperature of the monolayer (we set it to have the unit of thermal energy $k_B T$ ); $a_0 = A_{tot}/(NA_0)$ and $p_0 = -\gamma/(2\beta\sqrt{A_0})$ are the nondimensionalized preferred area and perimeter, respectively (it is easy to see that $\sum a_i = Na_0$); $A_{tot}$ is the total available area of the monolayer; $u_0 = (a_0 - 1)^2 + kp_0^2$ is a systematic energy term independent of the cell shapes. $a_0$ represents the compression ratio of available area to the preferred area, and we call it the *isotropic compression index* that directly reflects the cell density in the monolayer under gap-free constraint. The smaller $a_0$ is, the bigger cell density will be. There are other options for nondimensionalization. Here, we use the one that is consistent with that used in the recent discovery of the new order parameter, the *targeted shape index* $p_0$, for the unjamming transition[41].

We apply Metropolis Monte Carlo method[58, 59] to employ the stochastic concept, and systematically explore the collective behaviors of the gap-free monolayer in parameter space of $(\alpha_A, a_0, p_0)$. Simulations are performed in a square space with periodic boundary conditions. We set $k = 0.5$, $A_0 = 36\mu m^2$ and $N = 36$ unless otherwise stated. It is known that a biologically less relevant coarsening phase exists in the negative $p_0$ regime[41, 44, 56]. Because our focus is the unjamming transition between the solid-like and the liquid-like phases, we only study positive $p_0$ in this work. In addition to changing values of $\alpha_A$ and $p_0$ directly, we can control $a_0$ by tuning the edge length of the square (consequently $A_{tot}$). Existing study[41] and our preliminary data (unpublished) suggest that the system's phase transition is insensitive to $k$. Therefore, parameters $(\alpha_A, a_0, p_0)$ largely determine the evolvement of the monolayer, providing a comprehensive description of the phase transitions of the system. In the extreme case of $\alpha_A = \infty$, our method simply converges to the ground state analysis. The presented results are all obtained after the monolayer reaches steady state (Supplementary Fig. 1).

**Packing density $a_0$ shifts the critical transition point along the $p_0$ dimension**. Bi and coworkers recently discovered that without changing cell density, a monolayer could get unjammed *via* enhanced cell-cell adhesive interactions (increasing $p_0$)[41]. Under many physiological conditions, although there is no gap in the monolayer, cell number in a confined space can still change through cell division, cell death, cell extrusion and even direct mechanical loads. We first investigate whether and how cell density ($a_0$) affects the monolayer's unjamming transition. Using Equation (3), we explore the steady state morphological properties of the monolayer at different choices of $a_0$ and $p_0$.

As illustrated in Fig. 1a, two distinct categories of monolayer morphologies can be obtained when modulating $(a_0, p_0)$. The left panel exhibits a honeycomb like amorphous layout with rounded cells, representing a stable solid-like phase of the monolayer. The right panel shows a highly disordered layout with strip looking cells, representing a confluent liquid-like phase of the monolayer. In recognition of the vast difference in their geometries, we choose the cells' Diameter-Width Aspect Ratios (denoted as $AR$) as a geometrical indicator to distinguish the two phases (similar results can also be obtained using other geometrical indicators, Supplementary Fig. 2). As shown in Fig. 2b, a monolayer in the solid-like phase contains rounded cells that have larger $1/AR$ with low level of deviations; whereas the cells within the liquid-like phase, the monolayer have elongated cells that have smaller $1/AR$ (often correlated with biologically polarized state[60]) with high level of geometrical deviations. And from the colored contour plot (we call it the phase diagram) of $\langle 1/AR \rangle$ (averaging over all cells in the monolayer) in the $(p_0, \sqrt{a_0})$ plane (Fig. 1c), the two phases become immediately evident that the upper-left yellow region is the rounded solid-like jammed phase and the lower-right blue region is the elongated liquid-like unjammed phase, and the unjamming transition is gradual (Fig. 1d) with features of second-order phase transition. An interesting observation is that the boundary separating those two phases is approximately a straight line (Fig. 1c, Supplementary Fig. 2 and Supplementary Fig. 3). Fitting of our simulation results indeed shows that this boundary represents a linear relation: $p_0' = p_0^* \sqrt{a_0'}$, where $p_0'$ and $a_0'$ are the critical values at the transition boundary; and the slope $p_0^* = 3.91 \pm 0.06$ is close to the previously discovered critical transition value for $p_0$ under no compression condition[41].

In fact, if we assume that the density increase leads to uniform geometrical change of all cells until reaching the unjamming boundary, we can generalize the $P_0/\sqrt{A_0} \leq p_0^*$ constraint for cells being in solid-like state[41] to the general form $P_0/\sqrt{A_{tot}/N} \leq p_0^*$, where $P_0 = -\gamma/2\beta$. Therefore, the new constraint becomes $P_0/\sqrt{A_0} \times \sqrt{N \times A_0/A_{tot}} = p_0/\sqrt{a_0} \leq$

$p_0^*$, and the new transition boundary is then $p_0' = p_0^*\sqrt{a_0'}$. Our result indicates that the cell density and cell-cell interaction both contribute to and are capable of unjamming the cellular monolayer. More importantly, the critical values of the *targeted shape index* $p_0'$ and the *isotropic compression index* $a_0'$ linearly correlate with each other.

**Overcrowding induces $\alpha_A$-independent unjamming transition**. One surprising observation from Fig. 1c is that, smaller $a_0$ (i.e. higher cell density with higher isotropic compression) lowers the critical value of $p_0$ for the unjamming transition to happen. More specifically, if we fix the $p_0$ value and increase the packing density (moving down along an imaginary vertical line in Fig. 1c), the overcrowded monolayer gets unjammed, which is counterintuitive, as many-body systems generally get jammed when density increases under isotropic compression. For example, experiments have shown that granular materials undergo jamming-unjamming cycles when applied cyclic isotropic compressing-uncompressing loads accordingly[61].

To investigate whether and how $\alpha_A$ influences the overcrowding induced unjamming transition, we explicitly constructed the phase diagrams in the $(\sqrt{a_0}, \alpha_A)$ plane at $p_0$ values of 2.0, 3.0 and 4.0 (Fig. 2a-c). These results show that, when $\alpha_A$ is sufficiently high, increasing cell density (decreasing $a_0$) is consistently able to drive the monolayer to liquid-like state with elongated cell geometries. And for larger targeted shape index $p_0$, the unjamming transition happens earlier at bigger $a_0$. Fig. 2d-f exhibit the cross-section curves of the corresponding contour plots at various $\alpha_A$ values (1.6, 6.3, 11.0 and $\infty$). It is clear that decreasing $a_0$ induces a continuous yet dramatic unjamming transition; at the same time, as $\alpha_A$ increases and the system approaches to the ground state, the cross-section curves converge to sigmoidal shapes that depend on $p_0$.

From the phase diagrams at different $p_0$ values, the transition boundaries all seem to be vertical, suggesting that the overcrowding induced unjamming transition is independent of $\alpha_A$ as long as $\alpha_A$ is sufficiently large. To verify this, we analyze the relationship between $\sqrt{a_0}$ and $p_0$ at different $\alpha_A$ values and summarize the results in Fig. 2g. Different colors indicate different $\alpha_A$ values and each dot marks a pair of critical values along the corresponding phase boundary. All dots converge to the above identified linear line $p_0' = p_0^* \sqrt{a_0'}$, which overlaps with the transition boundary at the ground state (the dashed line). Therefore, overcrowding can drive the unjamming transition independent on the $\alpha_A$. Because $\alpha_A$ represents the combined measure of the mechanical stiffness and the relative scale of the uncoordinated active mechanical agitations (like the effective "temperature" of the system), our results suggest that, for different cell types with a broad range of mechanical stiffness and vastly different intracellular active forces, isotropic compression can always drive the cellular monolayer to the liquid-like collectively motile phase. And when the cells' *targeted shape index* $p_0$ is close to $p_0^*$, a moderate level compression is already sufficient to drive the unjamming transition.

It is worth noting that under extremely active force condition ($\alpha_A \to 0$), the monolayer's $\langle 1/AR \rangle$ value "recovers" to a relatively high level ($\sim 0.59$) comparing to the blue region ($\sim 0.2$). This does not mean that higher effective "temperature" heats the monolayer back to a more solid-like phase. In fact, small $\alpha_A$ results in low energy cost for any morphological changes. Therefore, the cells can "freely" explore a larger morphological space (rounded and elongated), so that $\langle 1/AR \rangle$ becomes larger than the liquid-like phase where cells are restricted to the elongated geometries (see SI). Future attention is needed to investigate whether the monolayer may reside in a gas-like phase under such condition and become highly spreadable.

**Free energy landscape of the T1 topological transition.** Long-range cell migration within a monolayer can be decomposed into a sequence of neighbor exchange events through cell intercalation processes known as the T1 topological transitions. As illustrated in Fig. 3a, a T1 transition involves 4 interconnected cells who switch their neighboring topology by first shrinking the inside edge between cells A and C (the pre-transition configuration) to a point shared by all 4 cells (the transition configuration), and then establishing a new inside edge between cells B and D (the post-transition configuration). From a standpoint of chemical physics, this T1 topological transition is reversible local event, and the free energy barrier of the transition directly determines the dynamic rate for cells to rearrange within the monolayer, which is therefore connected to the monolayer's overall motility[62]. However, existing evaluations of the barrier only take into account the mechanical energy difference between the transition configuration and the pre-transition configuration by modulating the length of the inside edge from a geometrical ensemble of the 4-cell groups[41, 44, 56]. This approach obtains the ensemble average of the mechanical energy contribution to the barrier, instead of the free energy barrier itself. To gain comprehensive insights towards the cellular rearrangements of the monolayer, we study from a statistical mechanics aspect to construct the free energy landscapes of the T1 topological transition during the unjamming process.

To construct the free energy landscape, the most straightforward way is to compute the probabilities of the inside edge's length by sampling a large quantity of the configurations of a group of 4 cells undergoing the T1 topological transition. Although this is technically not difficult, we employ a quicker method. Because any of the edges in the monolayer is the inside edge of certain 4 interconnected cells and it is equally possible to be in the pre- and post-transition configurations under no geometrical constraints, the steady state edge length distribution within the monolayer essentially resembles the inside edge length distribution of any given interconnected 4-cell group during the transition. Therefore, we can collect the steady state statistics of the edge length within the monolayer to construct the free energy landscape. Figure 3b-g shows a set of free energy landscapes of the T1 topological transition during the unjamming process driven by increasing the *isotropic compression index* $a_0$. The x-axis is the nondimensionalized edge length $d = D/\sqrt{A_0}$, where $D$ is the actual length of an edge.

Our results explicitly show the evolvement of the free energy landscape during the unjamming process from a single-barrier "W" shape to a double-barrier "M" shape through a "U" shape where the unjamming transition occurs. Starting from large $a_0$ where the monolayer is in the solid-like phase ($\sqrt{a_0} = 1.1$ and $1.0$), the energy landscape exhibits a "W" shape with two minimums at the pre- and post-transition configurations and a peak value at the transition configuration, indicating that cellular rearrangement in this phase requires non-*zero* "activation" energy. When cell density increases, the barrier becomes lower and eventually reaches *zero* ($\sqrt{a_0} = 0.85$). Similar barrier diminishing process has also been reported for the mechanical energy evaluation during the $p_0$ driven unjamming process[41]. After reaching the "U" shape, further decreasing $a_0$ results in a dip at the transition configuration where the system reaches its global minimum free energy state. At the same time, two new barriers emerge, leading to two local minima at the pre-/post-transition configurations ($\sqrt{a_0} = 0.8, 0.7$ and $0.6$). This double-barrier "M" shape has not yet been reported and directly reflects the elongated cell geometry where short and long edges coexist.

From the obtained free energy landscapes, we can directly measure the energy difference between the transition configuration and the local minima at the pre-/post-transition configurations, $\Delta F$ (Fig. 3b). When the *isotropic compression index* $a_0$ approaches the critical value to unjam the monolayer, $\Delta F$ linearly decreases from positive to *zero* (Fig. 4). After the monolayer gets unjammed, $\Delta F$ becomes negative and continues linearly decreasing with decreasing $a_0$ (Fig. 4). At different $\alpha_A$, all linear $\Delta F \sim a_0$ curves cross the same point ($\Delta F = 0$, $a_0 = a_0'$) but with different slopes (Fig. 4a), further confirming that the unjamming transition boundary is independent of $\alpha_A$ (i.e. cell stiffness and uncoordinated active forces). And at different $p_0$, on the other hand, the linear $\Delta F \sim a_0$ curves are shifted so that the transition boundary depends on $p_0$ (Fig. 4b).

Intriguingly, our results revealed two new free energy barriers for the T1 topological transition in the liquid-like phase: as the monolayer being unjammed, the original single energy barrier (peak at the transition configuration) is replaced by two barriers whose peaks locate at two sides of the transition configuration. Therefore, although $\Delta F$ becomes negative, the 4 participating cells still need to cross these new free energy barriers to rearrange. We believe that it is these two energy barriers that determine the cell rearrangement dynamics in the liquid-like phase. The detailed description of the free energy landscape provides a direct way to unbiasedly investigate collective cell motility and its connection to the monolayer morphology, which we will explore in future studies.

**Distributed-to-disordered geometrical transition during monolayer unjamming.** Previous studies have nicely demonstrated the geometrical distribution of the cells inside the monolayer and have shown a variety of coexisting cell shapes[13, 44, 46]. Here, we investigate whether and how the cell geometry evolves with the unjamming transition.

Figure 5 summarizes a collection of scatter plots of cell geometry. In each elementary panel, the x- and the y-axes are the square root of the area $\sqrt{a_i}$ and the perimeter $p_i$ of individual cells, respectively; each circle represents $(\sqrt{a_i}, p_i)$ of one cell in the monolayer at steady state with the designated parameters (see below). Different colors are used to indicate cells with different number of edges. Each row of panels has the same $a_0$ with descending values from top to bottom (1.5, 1.0, 0.6 and 0.3); each column of the panels has the same $p_0$ with ascending values from left to right (3, 4 and 5). With reference to Fig. 1c, the top left panels in Fig. 5 are in the solid-like phase and the bottom right ones in the liquid-like phase. This presentation allows us to quickly identify three distinct categories of morphological characteristics of the generally disordered monolayer: in the solid-like phase, the cells' $(\sqrt{a_i}, p_i)$ distribute along a linear line with correlation coefficients close to 1, and get compacted when the monolayer approaching the transition boundary; at the transition boundary, all circles collapse into a very narrow region around the critical transition point $(\sqrt{a_0'}, p_0')$ with correlation coefficients close to $-1$; and in the liquid-like phase, $(\sqrt{a_i}, p_i)$ of different cells become disordered without significant correlation. And cells in the solid-like phase always have $p_i > p_0$, whereas cells in the liquid-like phase always have $p_i < p_0$. Our results explicitly indicate that the unjamming transition coincides with a geometrical transition during which the cells' shapes within the monolayer transit from a distributed state to a disordered state.

In the solid-like phase, the cells' $(\sqrt{a_i}, p_i)$ exhibit near perfect linear correlations (upper left panels in Fig. 5 and Supplementary Fig. 5), suggesting isotropic and homogeneous deformation within the monolayer. The slope of this linear relation in the solid-like phase ranges from 2.6 to 3.1, which is significantly smaller than the geometrical constraints of pentagon (3.812) and hexagon (3.722). Closer inspection of the scatter plot reveals that, although cell shapes are distributed along a linear line, cells with smaller $(\sqrt{a_i}, p_i)$ tend to have fewer edges. In other words, triangles (if exist) would mostly locate in the bottom left, and when cell size increases, cells progressively become quadrangles, pentagons, hexagons and so on. These different polygons are like quantized geometrical states for cells to occupy when their sizes change (Fig. 6a). In Fig. 6b, we add the geometrical constraint lines, $p_i = c_n\sqrt{a_i}$, for different cell shapes, where $c_n$ is the geometrical constraint constant for a polygon with $n$ edges (distinguished using different colors). $c_n$ has values of 3.81, 3.72 and 3.67 for $n = 5$, 6 and 7, respectively. And every $(\sqrt{a_i}, p_i)$ circle has to stay above the line with the same color that is the constraint line according to its edge number. Thus, the $\sqrt{a_i} \sim p_i$ geometrical space is separated into different domains with different number of edges. It is clear that cells drift to domains with more edges when sizes increase, leading to the flattening of the distributed $(\sqrt{a_i}, p_i)$ scatter plot with a slope smaller than any of the individual geometrical constraint lines.

This trend of cell shape drifting provides a novel explanation for the unjamming transition. Completion of each T1 topological transition (Fig. 3a) causes cells A and C losing 1 neighbor and dropping to one-level lower quantized states, and cells B and D gaining 1 neighbor and jumping to one-level higher quantized states. The free energy barrier then corresponds to the resistance of jumping between adjacent quantized geometrical states. For large $a_0$ and small $p_0$, (Fig. 5 top left panels and Fig. 6ab) different quantized geometrical states are separated with little overlaps, so it is relatively difficult to jump between the states. When $a_0$ decreases (overcrowding) or $p_0$ increases (enhanced cell-cell adhesion), the distributed range gets compacted so that different quantized geometrical states merge closer with larger overlaps. And at the unjamming transition boundary, all circles get condensed to a very narrow and overlapped region (Fig. 5 panels circled by dashed line), indicating that cells may be able to change their number of neighbors more freely. Once the monolayer gets unjammed, the geometrical constraints, although still need to be satisfied, no longer dominate the cell shapes and the number of cell edges does not longer correlated with the size of the cell (Fig. 6cd).

**Discussion**

How crowdedly packed cells get unjammed and gain collective motility in tissues is an important open question. Although it is known that increasing cell density generally jams many-cell systems[27, 30, 31], in this communication, we report counterintuitive discoveries that increased cell density causes isotropic compression that unjams gap-free monolayers. We show that the monolayer, described by the vertex model, has an unjamming transition boundary determined by the *isotropic compression index* $a_0$ and the *target shape index* $p_0$: $p_0' = p_0^* \sqrt{a_0'}$, where $p_0^*$ is the critical transition value of $p_0$ under no compression condition. This overcrowding induced unjamming transition is also found largely independent on the uncorrelated mechanical agitations and the cell stiffness of the cells. Recently, it has been reported that mechanical compression unjams the human bronchial epithelial monolayer[63] by inducing elevated expression of folliculin, a major adhesive junction protein that directly modulates $p_0$ [42]. Here, our results suggest an alternative robust mechanism that the monolayers may spontaneously get unjammed through cell division and growth in confined spaces or by external isotropic compression.

Unlike the existing attempts that focus on either the ground state mechanical energy of the monolayer vertex[44, 56], or the average mechanical energy over an ensemble of metastable monolayer configurations[41], we directly construct the free energy landscape of the T1 topological rearrangements. Our analyses not only confirm the criticality of the overcrowding induced unjamming transition, but also explicitly show that the jammed and unjammed monolayers have distinct free energy landscapes: jammed phase has a "W" shaped landscape with single barrier at the topological transition configuration; whereas the unjammed phase possesses a "M" shaped landscape with two barriers separating the transition configuration from the pre- and post-transition configurations. During the unjamming transition, the single barrier diminishes and the two new barriers emerge. These results provide new and important insights on the dynamics of cellular rearrangements within the monolayer that may be considered as a morphological reaction process governed by the free energy landscape.

How does overcrowding unjam the gap-free monolayer? We believe this relates to the mechanical yielding of the monolayer. In fact, we believe that the recently discovered adhesion induced unjamming also reflects such mechanical yielding. Due to the gap-free constraint, the monolayer can be treated as a continuous material. On one hand, overcrowding can compress the monolayer isotropically: because the cells are soft and highly deformable, such isotropic compression quickly reaches the yielding point of the monolayer, leading to the unjammed liquid-like behaviors of the cells with highly disordered geometrical relations (Fig. 5 and 6); and the strong anti-correlation (with correlation coefficient nearly $-1$) between the cell's perimeter and area at the phase transition boundary indicates the onset of mechanical yielding that cells are isotropically compressed (smaller area) when they start to elongate (bigger perimeter). On the other hand, increasing the adhesion strength decreases the yield stress of the monolayer and can unjam the monolayer without additional mechanical compression. Our analyses rigorously confirm that such transition does not depend on uncoordinated active mechanical agitations and the overall cell stiffness, and further demonstrate the existence of a linear phase transition boundary $p_0' = p_0^* \sqrt{a_0'}$.

The discussed overcrowding and adhesion induced unjamming of gap-free monolayers differs from the unjamming processes of isolated many-cell systems. In isolated many-cell systems, each cell is intrinsically motile (i.e. self-propelled). Increased density and adhesion therefore enhance the probability of viscous cell-cell interactions, leading to locally synchronized motile groups and globally reduced mobility (jamming, illustrated in Fig. 7 as the packing process)[5, 27, 38]. In a gap-free monolayer, however, the mechanical integrity starts dominating the unjamming transition: increased density and adhesion unjam the monolayer mechanically by either compressing the monolayer beyond the yield stress or reducing the yield stress biologically (illustrated in Fig. 7 as the compressing process). This is in analogy to the yielding related jamming-unjamming transition of 2-dimensional metallic glasses[64].

Typical *in vitro* setup of monolayer experiments allows cells to increase or decrease their height when getting compressed within the 2-dimensional plane[40] and it is also discovered that extrusion occurs in epithelia to maintain homeostatic cell numbers[45, 65], which may help adjusting $A_0$ to maintain the *isotropic compression index* $a_0$ around 1. However, under many physiological conditions, such as solid tumor growth and embryo development, the available space is often limited that cells are confined/compressed by the surrounding tissue cells as well as the extracellular matrix[31, 66, 67] and the basement membrane[68, 69]. Thus, the overcrowding situation discussed in this communication becomes important and provides helpful insights on understanding the enhanced mobility of tumor and embryonic cells after reaching certain critical stages. From a mechano-biological coupling point of view, the biological cues induce intracellular biomechanical actions that may be in charge of navigating the migrations of the cellular assemblies; at the same time, the overcrowding by cell growth and division, together with the enhanced cell-cell adhesion, may then be in charge of removing the barrier of morphological rearrangements of the cells that enables and enhances the collective migrations of the entire cellular assemblies.

# Figures

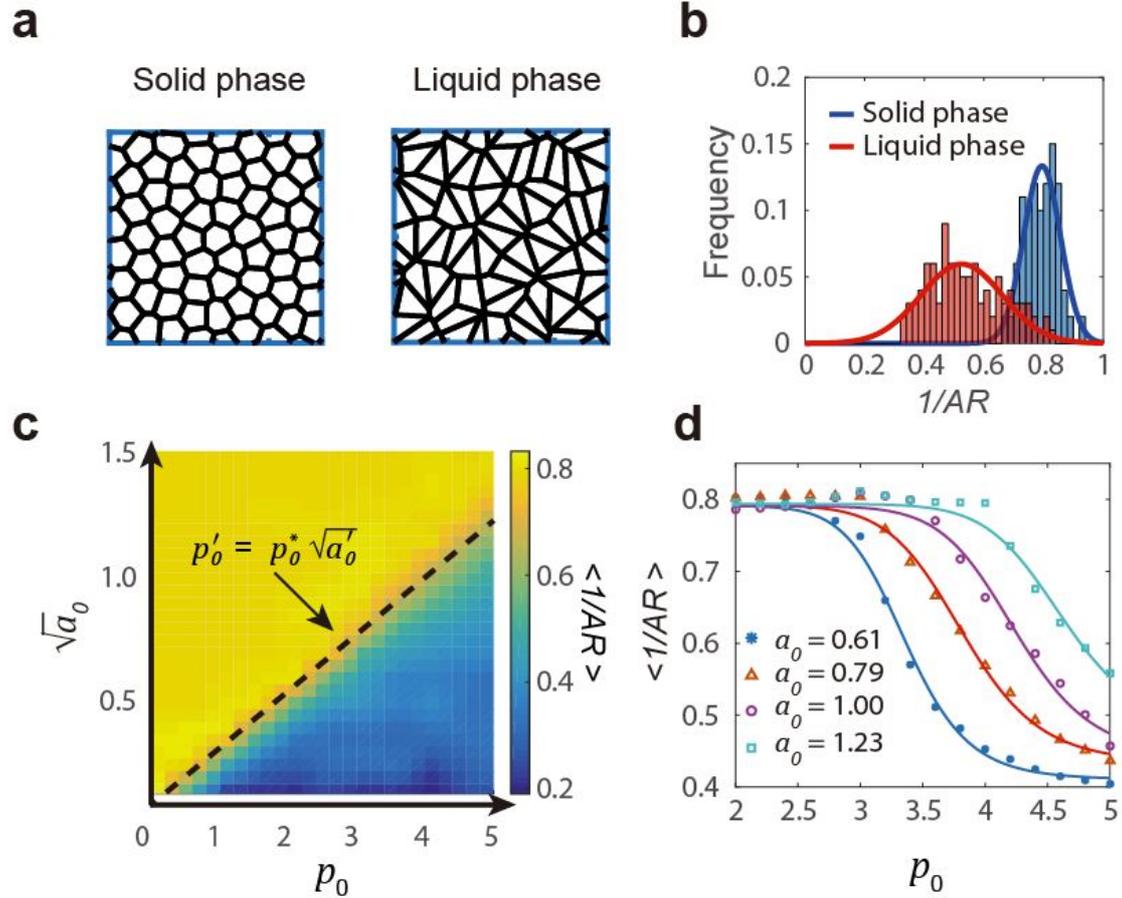

**Figure 1 | Overcrowding induced phase transition on a gap-free cell monolayer.** (**a**) Examples of cell monolayer's morphological characteristics and their cells' shapes: $a_0 = 1, p_0 = 3.5$, solid-like phase (left) and $a_0 = 1, p_0 = 4.5$, liquid-like phase (right), both obtained using $N = 100$, $k = 0.5$, $\alpha_A = 1000$. (**b**) Normalized distributions of $1/AR$ for the two phases shown in (**a**), and the curves are obtained from Gaussian fittings to show differences between their mean and standard deviation. (**c**) Phase diagram using $\langle 1/AR \rangle$ in the $p_0 \sim \sqrt{a_0}$ space, and the dashed linear line marks the boundary separating the solid-like (upper left) and the liquid-like (bottom right) phases. (**d**) $\langle 1/AR \rangle$ changes as function of $p_0$ at different $a_0$ values. Curves are for guidance of eyes and are fitted using Hill equations.

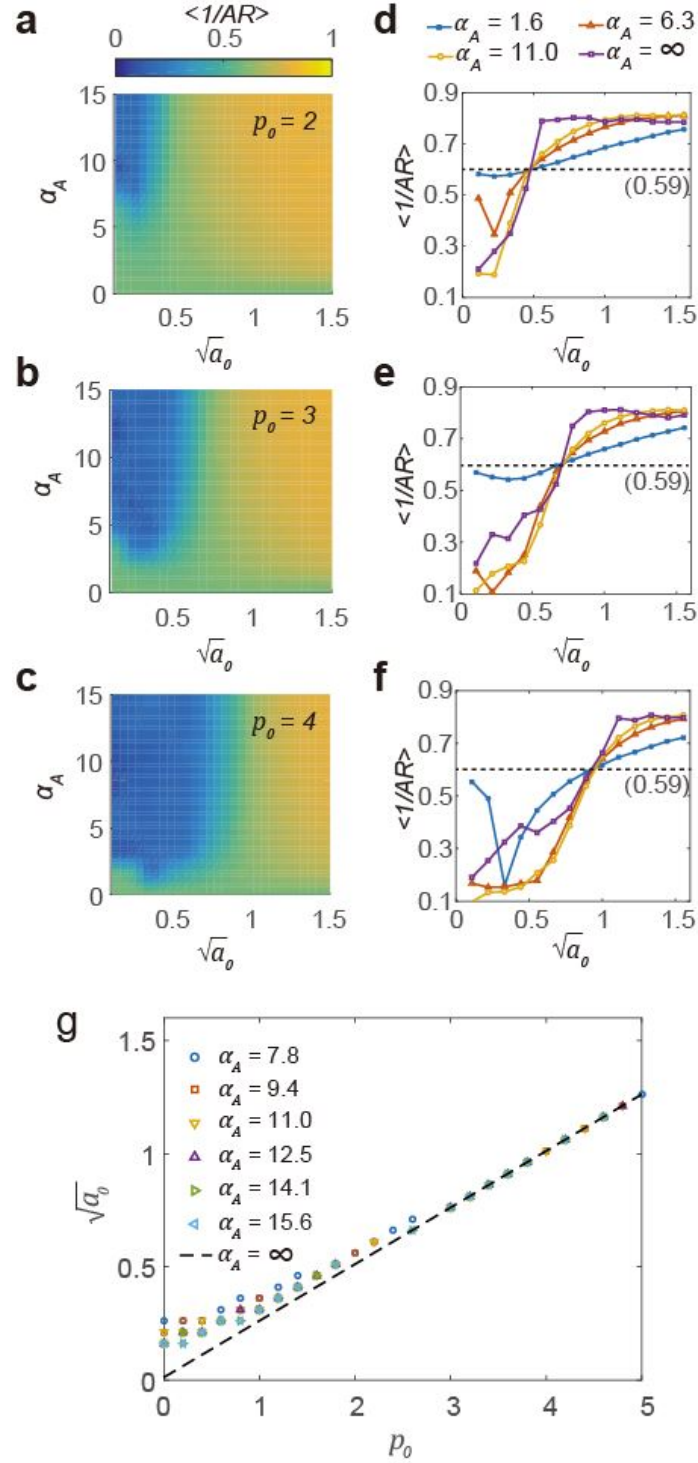

**Figure 2 | Overcrowding induced phase transition is independent on $\alpha_A$.** (**a-c**) Phase diagrams in the $\sqrt{a_0} \sim \alpha_A$ space at $p_0 = 2$ (**a**), $p_0 = 3$ (**b**) and $p_0 = 4$ (**c**). Colors represents the values of $\langle 1/AR \rangle$. (**d-f**) $\langle 1/AR \rangle - \sqrt{a_0}$ curves at different $\alpha_A$ values from corresponding phase diagrams shown in (**a-c**). The dotted lines are $\alpha_A = 0$ curves where $\langle 1/AR \rangle = 0.59$. (**g**) The $p_0' \sim \sqrt{a_0'}$ phase transition boundaries at different $\alpha_A$ values indicated using different colors. All dots converge nicely to the ground state phase transition boundary shown as the dashed line.

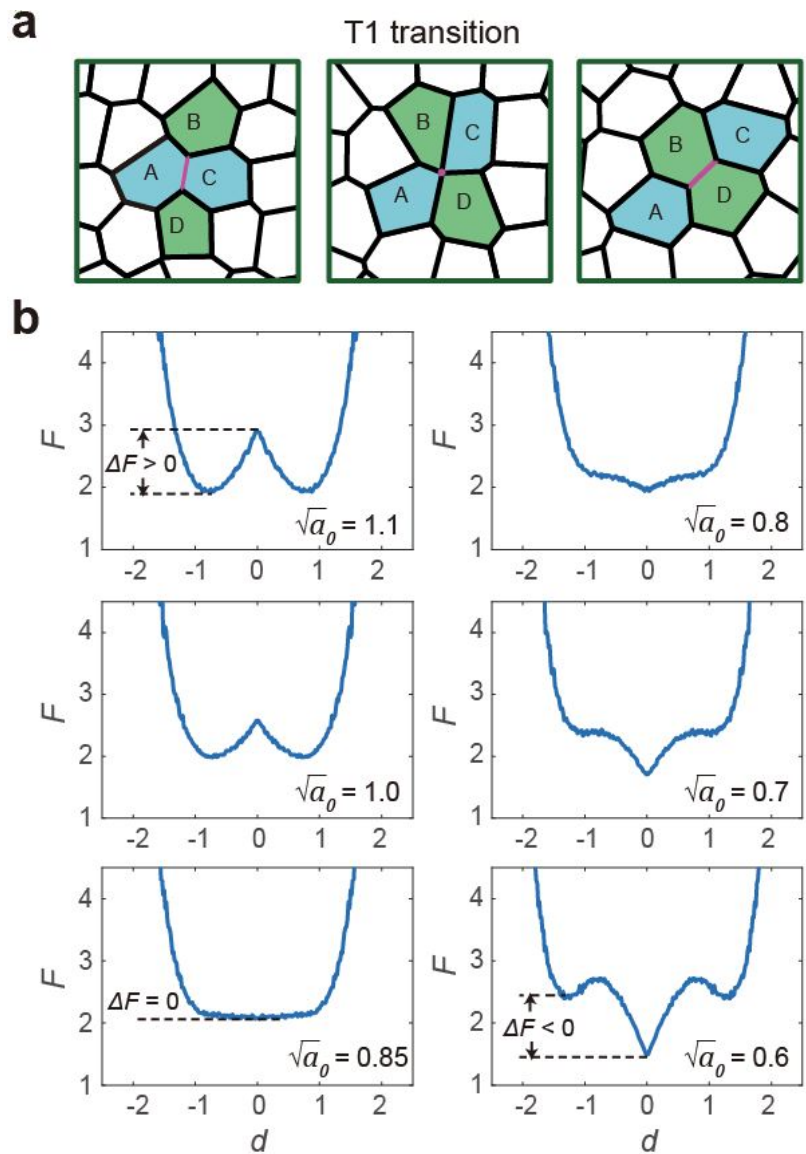

**Figure 3 | Free energy landscapes for the T1 topological transition during the overcrowding induced unjamming process.** (**a**) A typical T1 topological transition process: the pre-transition configuration (left), the transition configuration (middle) and the post-transition configuration (right). (**b**) The free energy landscapes at $p_0 = 3.5$, $\alpha_A = 6.3$ and different $\sqrt{a_0}$. For $\sqrt{a_0} = 1.1$ and $1.0$, the landscapes have a "W" shape with single barrier at the transition configuration. At $\sqrt{a_0} = 0.85$, the monolayer reaches the critical unjamming transition point without barrier. For $\sqrt{a_0} = 0.8$, $0.7$ and $0.6$, the landscapes have an "M" shape with two barriers in the pre- and post-transition regions. During the unjamming process, the free energy difference $\Delta F$ decreases and becomes negative. Each free energy landscape is obtained using $200$ snapshots of the monolayers at steady state.

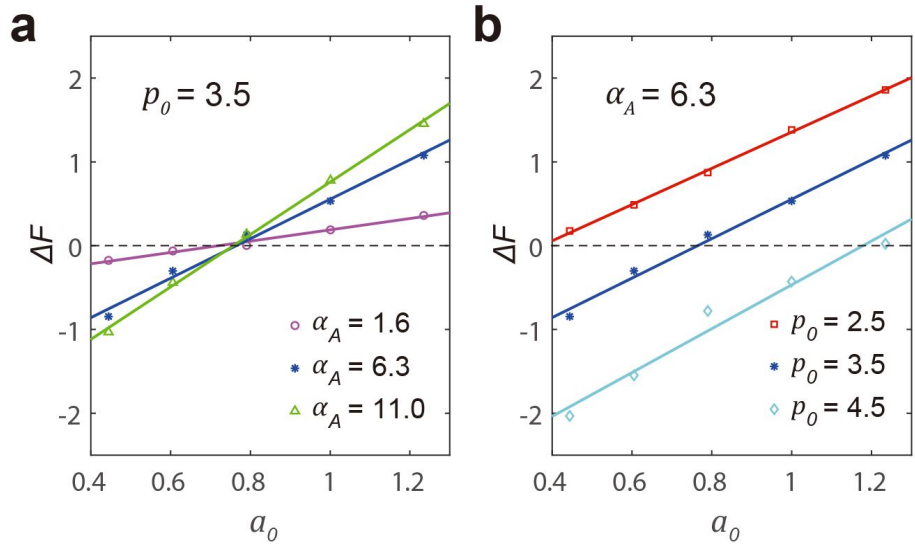

**Figure 4 | $\Delta F$ decreases linearly with decreasing $a_0$.** (**a**) At different $\alpha_A$ values, all $\Delta F - a_0$ curves cross the same phase transition point $(a_0', \Delta F = 0)$ with different slopes, indicating that the location of the phase transition boundary is independent of $\alpha_A$. (**b**) When $p_0$ is changed, the $\Delta F - a_0$ curves are shifted and unjamming transition occurs at different $a_0'$ values.

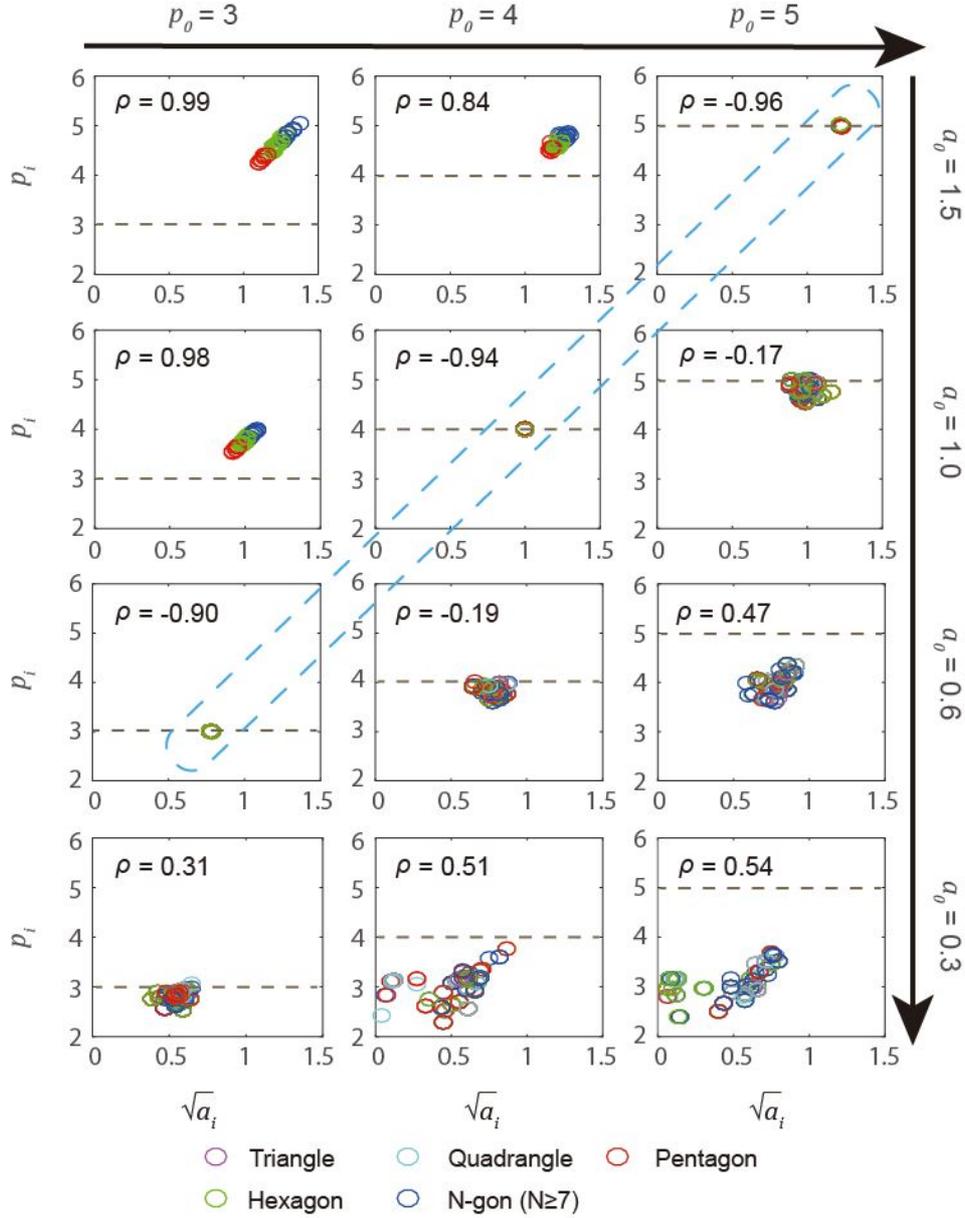

**Figure 5 |** $(\sqrt{a_i}, p_i)$ **scatter plots in the** $p_0 \sim \sqrt{a_0}$ **space ($\alpha_A = 1000$).** Each panel shows data from 100 snapshots of the monolayer with designated parameters (3600 data points). Magenta, cyan, red, green and blue circles represents cells with 3, 4, 5, 6, and ≥ 7 edges. Each column has the same $p_0$ value that increases from left to right; similarly, each row has the same $a_0$ value that decreases from top to bottom. The 3 upper left panels exhibit the distributed geometrical state of the monolayer in the solid-like phase. The 6 bottom right panels exhibit the disordered geometrical state of the monolayer in the liquid-like phase. The 3 panels circled by dashed line are the transition boundary where all the $(\sqrt{a_i}, p_i)$ circles get highly compacted. $\rho$ is the correlation coefficient between $\sqrt{a_i}$ and $p_i$ for each panel.

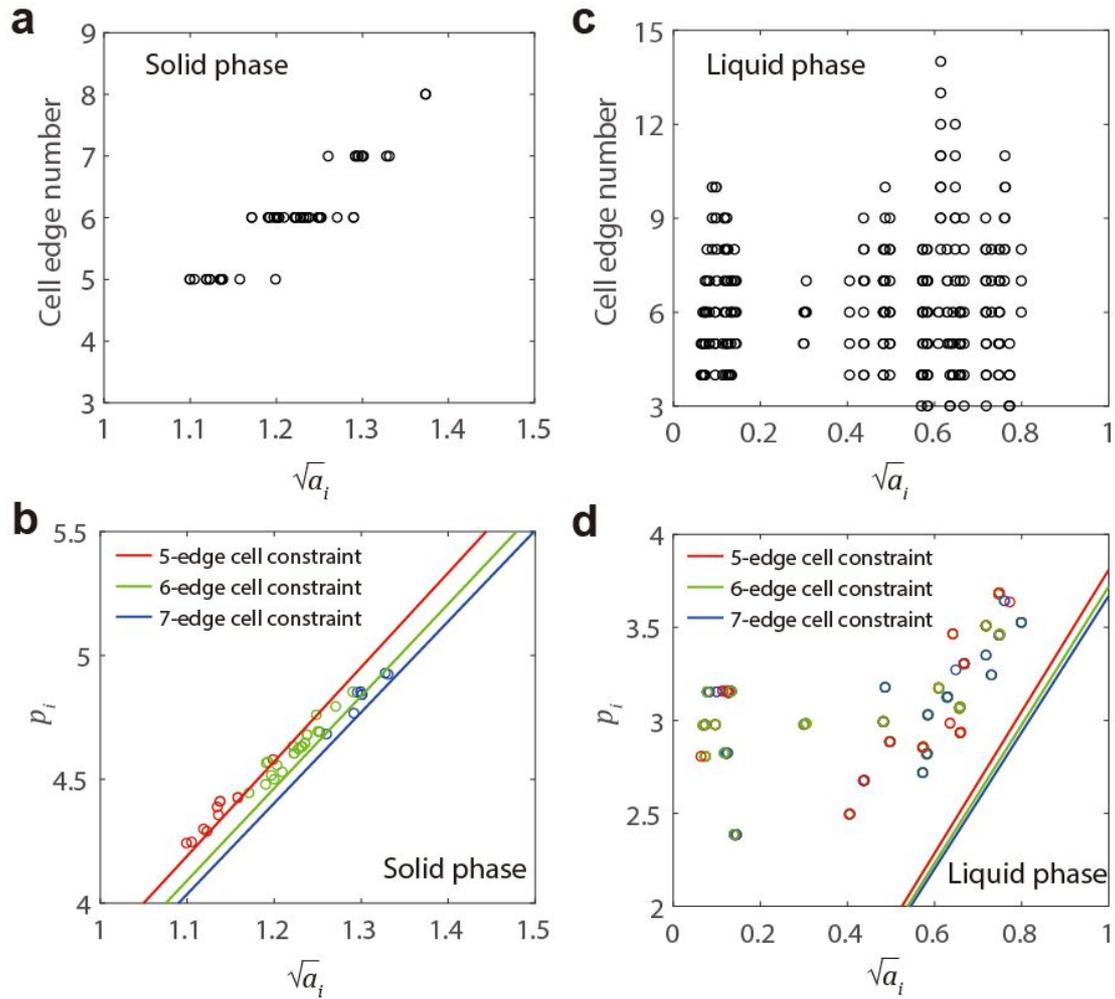

**Figure 6 | Cell edge numbers of two typical cases as function of $\sqrt{a_0}$ ($\alpha_A = 1000$).** In the solid-like phase, (**a**) smaller cells tend to have fewer edges; (**b**) different geometrical domains are separated by constraint lines and cells drift to domains with more edges when getting bigger. In the liquid-like phase, (**c**) cell size does not correlate with the number of edges; (**d**) the geometrical constraint still holds but does not dominate the cell geometry.

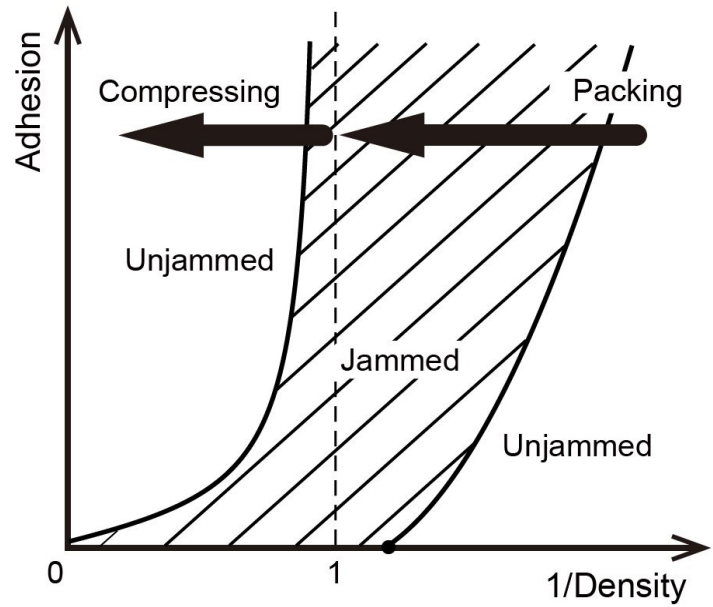

**Figure 7 | Schematic illustration of packing induced jamming and compressing induced unjamming processes.** The shadowed region is the jammed phase. For isolated many-cell systems, increasing cell density and cell-cell adhesion jams the monolayer by a "packing" process. For gap-free monolayer, increasing cell density and cell-cell adhesion unjams the monolayer by "compressing" the soft material to yield.